\documentclass[9pt,twocolumn,twoside]{osajnl}
\journal{ol} 
\setboolean{shortarticle}{true} 

\title{Infrared dielectric properties of low-stress silicon oxide}

\author[1,*]{Giuseppe Cataldo}
\author[1]{Edward J. Wollack}
\author[1]{Ari D. Brown}
\author[1]{Kevin H. Miller}

\affil[1]{NASA Goddard Space Flight Center, 8800 Greenbelt Road, Greenbelt, MD 20771, USA}
\affil[*]{Corresponding author: Giuseppe.Cataldo@NASA.gov}

\dates{Compiled \today}

\ociscodes{(160.4760) Optical properties; (310.3840) Materials and process characterization; (310.6188) Spectral properties; (310.6860) Thin films, optical properties.}

\doi{\url{http://dx.doi.org/10.1364/OL.41.001364}}

\begin{abstract}
Silicon oxide thin films play an important role in the realization of optical coatings and high-performance electrical circuits. Estimates of the dielectric function in the far- and mid-infrared regime are derived from the observed transmittance spectrum for a commonly employed low-stress silicon oxide formulation. The experimental, modeling, and numerical methods used to extract the dielectric function are presented.
\end{abstract}

\setboolean{displaycopyright}{true}

\begin{document}

\maketitle
\thispagestyle{fancy}
\ifthenelse{\boolean{shortarticle}}{\abscontent}{}

Silicon oxide (SiO$_{\mbox{\scriptsize x}}$) is widely employed as a dielectric medium due to its low loss, insulating properties, and general compatibility with optical coating and micro-fabrication processing~\cite{Kitamura,Lamb}. Thin silicon monoxide films have demonstrated acceptable dielectric performance for high-frequency applications, as a dielectric medium~\cite{Watkins}; however, the achievable loss tangent is dependent on the deposition rate, annealing and trace elemental constituents (H, C, OH, etc.) incorporated during deposition, and subsequent use. From this perspective SiO$_{1.5}$ is a preferable stoichiometric composition in order to reduce the number of free bonds and minimize the dielectric medium's absorption~\cite{Allam}. The strength and details of the infrared bands are of particular importance in determining the behavior of amorphous solids such as silicate glasses. In the far-infrared the absorption coefficient of glasses is largely featureless, scales as the square of frequency, and typically exceeds that of crystalline solid counterparts by an order of magnitude due to optical coupling to Debye-like and lattice modes~\cite{Strom}. These general features arise in glasses from spatial and temporal disorder broadening of the lattice absorption bands into a continuum. Here, the infrared properties of low-stress silicon oxide films are characterized and compared to materials reported in the literature.

\begin{figure}[htbp]
	\centering
	\includegraphics[width=0.45\textwidth]{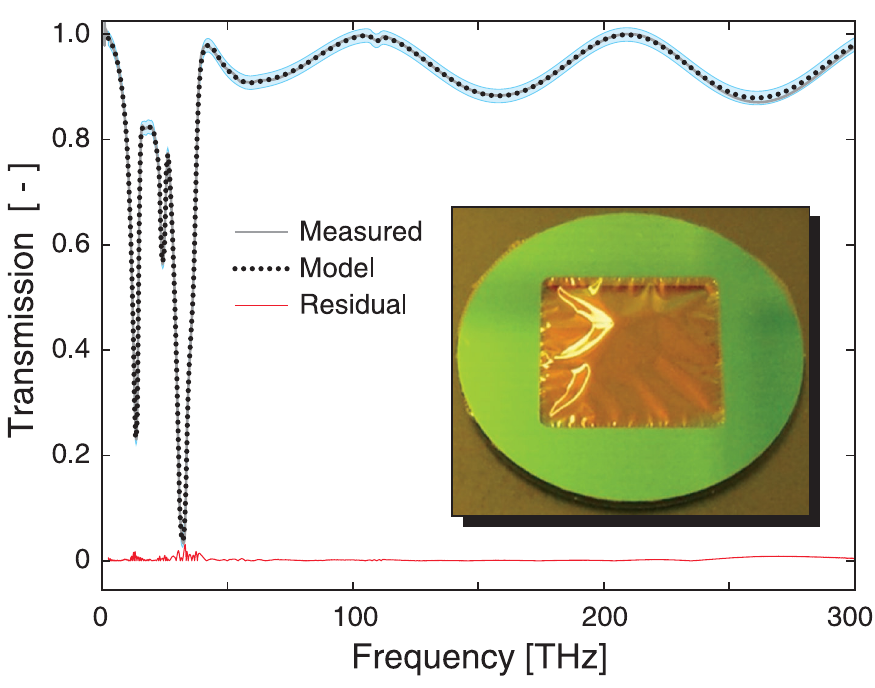}
	\caption{Room-temperature transmission of a silicon oxide sample: measured (grey), model (black dots), and residual (red). The shaded band's width delimits the estimated 3$\sigma$ measurement uncertainty.
	A 7.2~GHz (0.24~cm$^{-1}$) resolution is employed for the measurements and the model is smoothed to the same value. The insert shows a single test sample made of a silicon oxide membrane supported by a 12.75-mm-diameter silicon frame.}
	\label{fig:T_1um}
\end{figure}

The amorphous silicon oxide films were prepared by PECVD (plasma-enhanced chemical vapor deposition) on H-terminated 100-mm Si(001) substrates. A 1000-W microwave plasma with a 2:1 O$_2$:SiH$_4$ gas ratio at 3.4~mT was used to grow the silicon oxide and a 50-W RF bias was used to densify it. This process enabled low-compressive-stress (<~200~MPa) films critical for yielding free-standing membranes. The resulting samples are consistent with the Fourier transform spectrometer (FTS) throughput requirements for optical characterization and use as a microwave dielectric substrate~\cite{Brown}. Each silicon oxide membrane is a square 6.45~mm on a side and has a 12.75-mm-diameter Si frame. Each wafer contains 19 samples. See insert in Fig.~\ref{fig:T_1um} for a photo of a representative sample. 

The silicon-oxide-coated side of the wafers was lithographically patterned with a street cut mask to define the perimeter of the samples and the silicon oxide was reactive ion etched in a CF$_4$/CHF$_3$/Ar plasma. The thickness, 1.020~\textmu m, was determined with an~$\alpha$-SE spectroscopic ellipsometer using a calibrated SiO$_2$ thin-film reference standard. The wafers were then wax bonded to 100-mm Pyrex disks with Crystalbond-509.
A photoresist mask and deep reactive ion etching were used to define the silicon oxide windows and frames with the Bosch process. When the silicon oxide layer was exposed near the wafer edge, the RF power used to etch the silicon was decreased from 600~W to 400~W in order to minimize damage to the silicon oxide and prevent the wax from reflowing. After release in acetone, the intrinsic film stress of the silicon oxide induced corrugations along the membrane perimeter. Initial transmission measurements of the samples revealed the presence of a resonance at $\sim$~3600~cm$^{-1}$, which indicated the presence of --OH or water incorporated in the film~\cite{Noguchi}. The samples were annealed at $600^{\circ}$C for one hour in vacuum ($1\times10^{-8}$~Torr). The residual pressure of H$_2$ in the chamber increased fiftyfold momentarily  at $\sim400^{\circ}$C and then recovered to the background value. Annealing results in a more quartz-like structure with an increase in the static dielectric permittivity and an upward shift in the 10~\textmu m line~\cite{Garski}. 

The transmission of a silicon oxide membrane was characterized with a Bruker IFS 125 high-resolution FTS. The sample was placed at a focus ($\sim$~1\,mm diameter) in a f/6.5 focused beam geometry in order to mitigate the influence of the flatness variations. The data repeatability was verified by measuring at several positions. Three combinations of source, beamsplitter, and detector were implemented to span the frequency band 30--10,000~cm$^{-1}$.
In the far-infrared range a black polyethylene filter was used to limit the thermal heat load on the bolometer. A red blocking filter was used in the near-infrared range in conjunction with the tungsten source to achieve a more balanced single-beam intensity spectrum over the range of interest. Transmission data from the three ranges agreed to within 0.5\% in their regions of overlap. The three datasets were merged into one spectrum by using a weighted average (Fig.~\ref{fig:T_1um}).

\begin{figure}[!tp]
	\centering
	\includegraphics[width=0.44\textwidth]{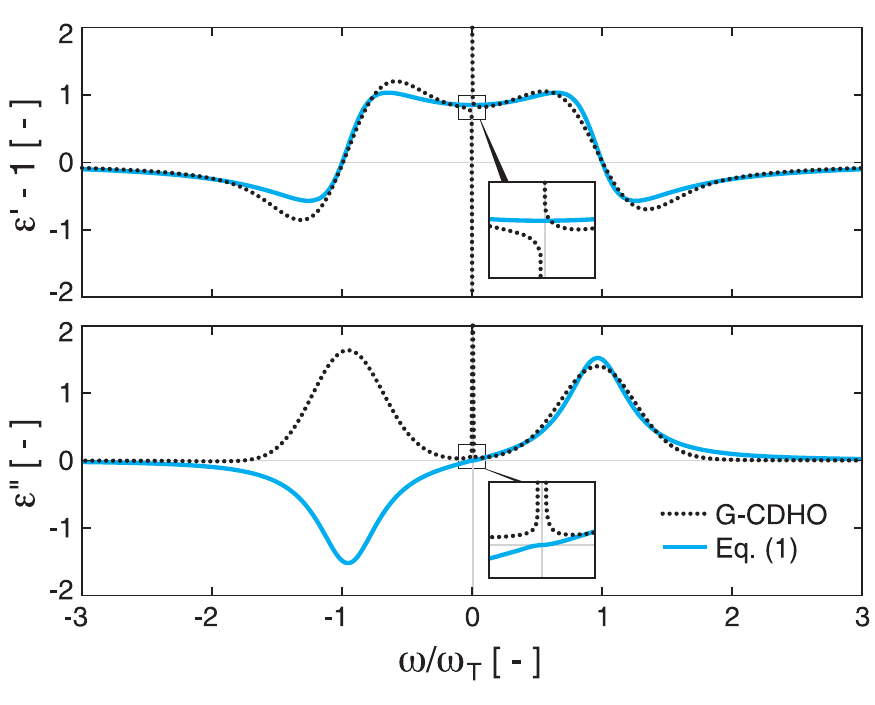}
	\caption{Real and imaginary parts of the G-CDHO and the Maxwell-Helmholtz-Drude dielectric functions. The following parameters were used to evaluate the G-CDHO profile~\cite{BB} (black dots): $f_j\omega_p^2 = \Delta\varepsilon' = 0.6$, $\Gamma = 0.34\,\omega_T$, $\sigma = 0.27\,\omega_T$. The G-CDHO profile does not meet the parity requirements and diverges to infinity as $\omega\rightarrow0$.
	The following parameters were used to evaluate the Maxwell-Helmholtz-Drude profile (\eqref{eq:eps}, blue solid): $\Delta \varepsilon' = 0.85$, $\varepsilon_\infty = 1$, $\Gamma = 0.20\,\omega_T$, $\alpha = 0.0001$. This meets the Kramers-Kronig causality criteria, i.e., the function is analytic over the upper part of the complex plane, it converges to a real constant for $\omega\rightarrow\infty$, and its real and imaginary components are even and odd functions respectively~\cite{Keefe}. As required for a passive medium, the imaginary component, $\varepsilon''$, is positive for $\omega>0$.}
	\label{fig:models}
\end{figure}

To interpret the observed response, the transmission spectrum was analyzed in terms of a classical dielectric function, which physically relates the driving excitation to the material's electromagnetic response. In the infrared, the dielectric function response is specified by a series of resonances and line shape profiles. Commonly used line shape profiles include the Lorentzian, the Gaussian and their convolution (the Voigt)~\cite{Armstrong}. These real-valued profiles are valid in the optical regime, where the frequency is approximately equal to the resonance frequency and is large compared to the line width $(\omega \approx \omega_{\mbox{{\tiny\it T}}} \gg \Gamma)$. With the appropriate symmetrizations, these dielectric functions satisfy the Kramers-Kronig relations~\cite{Keefe}.
In the infrared, the Gaussian-broadened complex damped harmonic oscillator (G-CDHO) has greater fidelity in reproducing the properties of amorphous dielectric solids~\cite{Efimov,BB} and metals~\cite{Rakic}. In the latter case, a Drude term (resonance at zero frequency) is added to describe the finite conductivity in the medium. The formulation found in~\cite{BB} is plotted in Fig.~\ref{fig:models}. This convolved profile better approximates the line shape but fails in reproducing the profile's wings in detail for amorphous materials. In addition, the resulting relative complex permittivity, $\hat{\varepsilon}$, does not meet the parity requirements of the Kramers-Kronig relations and diverges to infinity as $\omega\rightarrow0$~\cite{DeSousa}. Mathematically, this occurs because $\hat{\varepsilon}$ is proportional to $(\omega^2-i\omega\Gamma)^{-1/2}$. The resulting profile has no adjustable parameters to specify the scaling and amplitude of the dielectric function as $\omega\rightarrow0$. A finite sum of oscillators with Gaussian weights~\cite{Wakino} does not suffer from this fundamental limitation.

Here, the classical Maxwell-Helmholtz-Drude dispersion model~\cite{Button} is used to reproduce the spectrum in Fig.~\ref{fig:T_1um}:
\begin{equation}
	\label{eq:eps}
	\hat{\varepsilon}(\omega) = \varepsilon_\infty+ \sum_{j=1}^M \frac{\Delta\varepsilon'_j\cdot\omega^2_{\mbox{{\tiny\it T}}_j}}{\omega^2_{\mbox{{\tiny\it T}}_j}-\omega^2-i\omega\Gamma'_j(\omega)}.
\end{equation}
In \eqref{eq:eps}, $\hat{\varepsilon}(\omega)$ is a frequency-dependent function of $4M+2$ degrees of freedom (DOF), which are as follows: the high-frequency relative complex permittivity, $\varepsilon_\infty = \varepsilon'_{M+1}$; the difference in real part of the relative dielectric constant between adjacent oscillators, $\Delta\varepsilon'_j=\varepsilon'_j-\varepsilon'_{j+1}>0$, which serves as a measure of the oscillator strength; the oscillator resonance frequency $\omega_{\mbox{{\tiny\it T}}_j}$; and the effective Lorentzian damping coefficient $\Gamma'_j$, for $j=1,...,M$ and $M$ the number of oscillators. The static value is $\hat{\varepsilon}(0) = \varepsilon_\infty + \sum_j \Delta\varepsilon'_j$. The following functional form is used to specify the damping:
\begin{equation}
	\Gamma'_j(\omega)=\Gamma_j\exp{\left[-\alpha_j\left(\frac{\omega^2_{\mbox{{\tiny\it T}}_j}-\omega^2}{\omega\Gamma_j}\right)^2\right]}
\label{eq:Gauss}
\end{equation}
where $\alpha_j$ allows interpolation between Lorentzian $(\alpha_j=0)$ and Gaussian $(\alpha_j>0)$ wings~\cite{Cataldo2012}. The functional form of~\eqref{eq:eps} satisfies the requirements of causality and passivity of the Kramers-Kronig relations~\cite{Keefe}, as shown in Fig.~\ref{fig:models}.

\begin{figure*}[!t]
	\centering
	\includegraphics[width=0.80\textwidth]{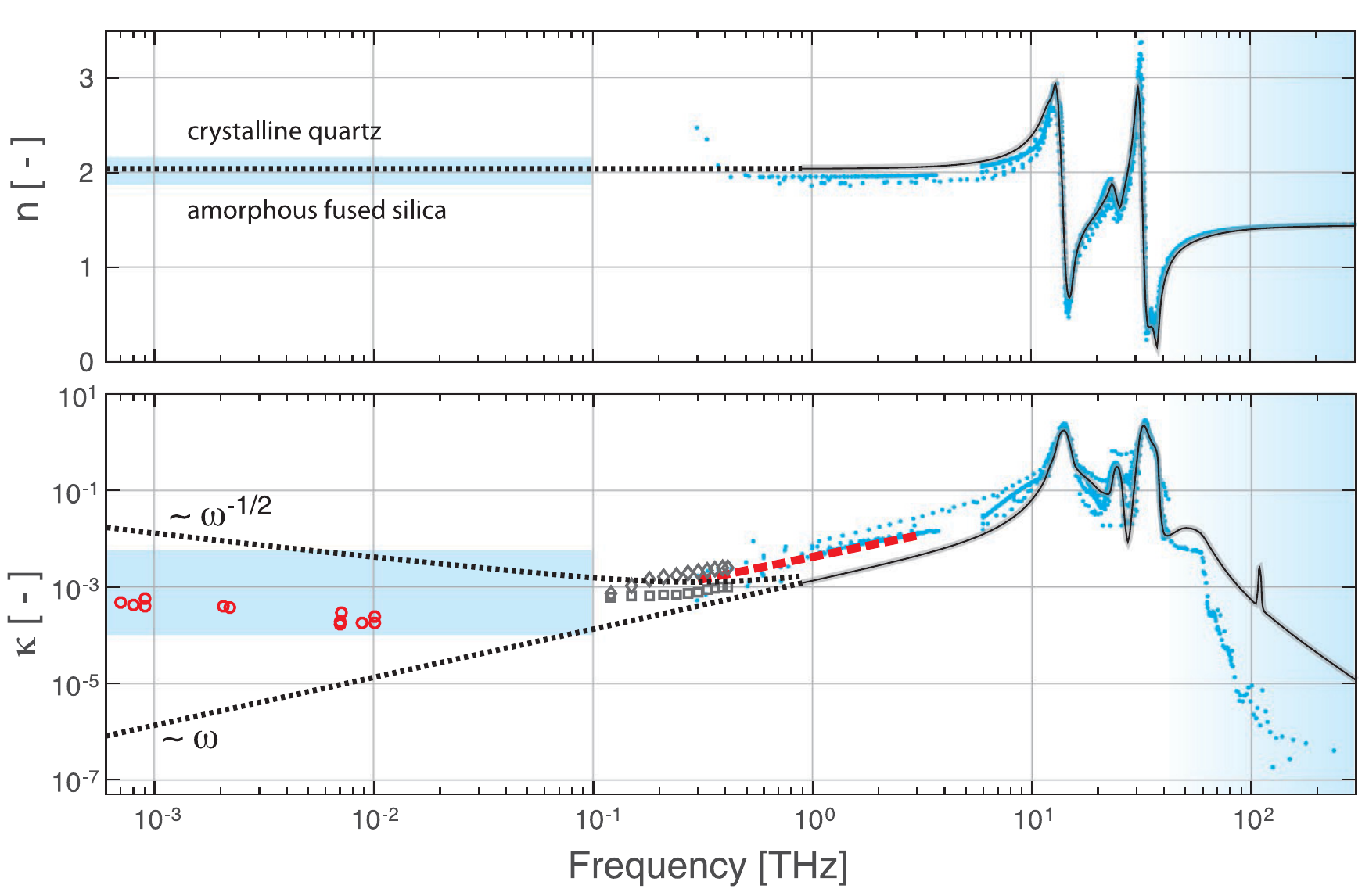}
	\caption{Real and imaginary components of the complex refractive index (black line) extracted from the measured transmittance. The grey shaded line is indicative of the $\sim$ 3\% error band propagated from the measured spectrum. Data from the literature~\cite[and references therein]{Kitamura} are overlaid for comparison (blue dots). Also included are data for fused quartz (red dashes)~\cite{Bagdade}, UV-grade fused silica (grey diamonds) and water-free fused silica (grey squares)~\cite{Afsar}. The shaded boxes bound the scatter of the refractive and absorption indices found in the microwave literature~\cite{Lamb} and the SiO$_{\mbox{\scriptsize x}}$ data (red circles) from~\cite{Kaiser}.
	In the lower panel, low-frequency dielectric function scaling laws are indicated by dotted lines. A divergence as $\omega^{-1/2}$ can occur in amorphous materials due to Debye relaxation losses~\cite{Ngai}.
	In non-conducting crystalline materials $\kappa$ tends to increase as $\omega$~\cite{Gurevich,Petzelt}.
	 The shaded gradient indicates the region where the form for $\Gamma'(\omega)$ adopted does not represent multiphonon processes.}
	\label{fig:nk_1um}
\end{figure*}

The impedance contrast between the sample and the vacuum forms a Fabry-Perot resonator as can be seen in Fig.~\ref{fig:T_1um}. The dielectric function, \eqref{eq:eps}, was used to model the measured power transmission as a function of frequency and sample thickness, $h$~\cite{BohrenHuffman}. A non-linear least-squares fit of the modeled transmission, $T$, to the laboratory data, $T_{\mbox{\tiny{FTS}}}$, in Fig.~\ref{fig:T_1um} was performed by solving the following minimization problem:
\begin{equation}
	\underset{\mbox{\scriptsize DOF}}{\mbox{min}} \, \chi^2 = \underset{\mbox{\scriptsize DOF}}{\mbox{min}} \sum^N_{k=1}{\left[ T(\hat{\varepsilon}(\omega_k),h) - T_{\mbox{\tiny{FTS}}_k} \right]^2},
	\label{eq:chi2}
\end{equation}
where $N$ is the number of data points. A sequential quadratic programming (SQP) algorithm was implemented with finite-difference computation of the Jacobian and Hessian matrices. 
Since the variables differ from each other by several orders of magnitude, a scaling of the variables and constraints was performed at each iteration to condition the problem and speed convergence.

The modeled results for the transmission are shown in Fig.~\ref{fig:T_1um} laid over the measured data. The observed peak residual in the transmission is less than 0.032 and the $3\sigma = 0.013$ uncertainty band indicated in Fig.~\ref{fig:T_1um} corresponds to the 99.7\% confidence level under the assumption that the errors are uniform. The systematic uncertainty in the spectrum has contributions arising from the change in illumination between the sample and reference positions and the calibration of the differing FTS configurations. In analyzing the data, it was noted that the maximum transmittance value was 1.0007. This can arise from internal reflections in the instrument ($\approx 0.005$ amplitude, $3$~cm$^{-1}$ fringe rate) leading to errors in the calibrated transmittance. To access the potential influence of these systematics, the spectrum was renormalized by 1.005 and reanalyzed. The dominant observed effect was a $\approx 1\%$ fractional change in the extracted $\varepsilon'_j$ in Table~\ref{tab:parameters}.

From a frequentist statistical perspective, 12 oscillators are identified with the vibrational modes found in the literature (Table~\ref{tab:parameters}). In principle, adding more resonators could enable smaller residual errors in the modeled transmittance. For example, when fitting the data in Fig.~\ref{fig:T_1um} with the G-CDHO model, 17 oscillators were identified with residuals $<2\%$. However, from a Bayesian standpoint~\cite{Kass} only 12 oscillators were statistically justified. Therefore, in interpreting the data through a model, overfitting the residuals between the model and the data presents a concern. The residual features can migrate to the absorption spectrum and give the impression that additional sharp transitions exist~\cite{Kischkat}.

The real and imaginary components, $n$ and $\kappa$, of the complex refractive index, $\hat{n}$, are shown in Fig.~\ref{fig:nk_1um} and are computed from:
\begin{equation}
	\hat{\varepsilon} = \varepsilon' + i\,\varepsilon'' = \hat{n}^2 = (n + i\,\kappa)^2 =  (n^2-\kappa^2)+i\;2n\kappa,
	\label{eq:nk}
\end{equation}
where a nonmagnetic permeability $\mu_r =1$ is adopted.
For the interpretation of the FTS data of low-loss materials, $\hat{n}$ is the preferred parameterization to prevent contamination of the imaginary component of the dielectric function~\cite{Afsar}. Figure~\ref{fig:nk_1um} shows a comparison of the values of $n$ and $\kappa$ with data from the literature~\cite[and references therein]{Kitamura}. This is augmented with data from~\cite{Lamb,Bagdade,Afsar,Kaiser,Gurevich,Petzelt}. The variability between the literature data is significant and can be traced to the different stoichiometric compositions, sample purity and experimental techniques; in some cases the description provided is inadequate to inform a detailed comparison.

\begin{table*}[!t]
  \centering
  \caption{Fit parameters. The high-frequency relative permittivity is $\varepsilon_\infty = \varepsilon'_{13} \approx 2.08212(2)$. The significant digits are provided for data reproducibility purposes. The values in parentheses indicate the uncertainties propagated from the transmission residuals.}
    \begin{tabular}{lccccll}
    \hline
     $j \ [-]$	& $\varepsilon'_j \ [-]$	& $\omega_{\mbox{{\tiny\it T}}_j}/2\pi \ [\mbox{THz}]$	& $k/2\pi \ [\mbox{cm}^{-1}]$	& $\Gamma_j/2\pi \ [\mbox{THz}]$	& $\alpha_j \ [-]$	& Mode \\
    \hline
      1    & 4.1728(2)7	& 12.685(4)3 	& 422.8(4)793	& 0.4415(4)3	& 0.0000(0)	& Si--Si ``breathing'' \cite{Bell,Tsu}\\
      2    & 3.869(8)87	& 13.4884(1) 	& 449.61(3)77	& 1.40(0)526	& 11.8(5)73	& Si--O rocking \cite{Bell,Tsu}\\
      3    & 3.6699(7)3	& 13.5791(2)	& 452.637(3)4	& 0.21372(7)	& 0.458(7)3	& Si--O rocking \cite{Bell,Tsu}\\
      4    & 3.0434(7)9	& 16.876(5)2	& 562.55(0)67	& 0.5242(3)5	& 0.755(5)3	& Si--O bending/stretching~\cite{Bell}\\
      5    & 2.9623(3)9	& 20.490(1)8	& 683.00(6)17	& 0.9784(5)8	& 1.698(0)4	& Si$_2$--Si--H$_2$ wagging (?) \cite{Theiss}\\
      6    & 2.88419(8)	& 24.4048(7)	& 813.495(7)0	& 0.3696(7)7	& 0.5127(2)	& Si--O bending \cite{Bell,Tsu}\\
      7    & 2.78409(5)	& 31.6703(3)	& 1055.677(7)	& 0.3377(4)3	& 0.2098(5)	& Si--O stretching \cite{Bell,Tsu}\\
      8    & 2.15241(9)	& 35.3860(7)	& 1179.53(5)6	& 0.4988(2)5	& 0.5228(4)	& Si--O--Si stretching \cite{Theiss}\\
      9   & 2.09731(9)	& 49.34(8)65	& 1644.9(5)53	& 2.447(7)82	& 0.000(0)		& O--H bending \cite{Miller}\\
      10 & 2.08997(7)	& 56.95(6)88	& 1898.(5)627	& 2.903(4)13	& 0.000(0)		& O--H bending \cite{Miller}\\
      11  & 2.08250(3)	& 58.44(9)17	& 1948.(3)059	& 0.773(9)31	& 0.000(0)	 	& Si--H$_{\mbox{\scriptsize{x}}}$ stretching \cite{Theiss}\\
      12  & 2.08232(4)	& 109.220(8)	& 3640.6(9)38	& 0.597(6)95	& 0.74(9)57	& O--H stretching \cite{Tsu}\\
      \hline
    \end{tabular}
  \label{tab:parameters}
\end{table*}

The measured FTS data (Fig.~\ref{fig:T_1um}) do not highly constrain the complex refractive index below 0.8~THz and above 100~THz. However, it is useful to consider the behavior of the dielectric function outside the range where the modeled data are valid in Fig.~\ref{fig:nk_1um}. For amorphous dielectric materials, in the limit of $\omega\rightarrow0$, a Debye term is typically required to account for the polarization relaxation mechanism or, more generally, a universal response associated with the presence of very-low-energy excitations in the medium~\cite{Ngai}. For $\omega$ greater than the highest infrared transverse optical mode, the absorption is dominated by multiphonon processes~\cite{Thomas, Sham}. This effect is not addressed in the model defined by \eqref{eq:eps} and can be seen as an excess in the profile wings above $\sim50$\,THz (see shaded region in Fig.~\ref{fig:nk_1um}). With spectral data spanning the transmission window, the approach described in~\cite{DeSousaB} can be employed to uniquely parameterize and represent this detail.

The properties of a low-stress silicon oxide film suitable for microwave and terahertz applications have been presented and contrasted with other variations reported in the literature. The symmetry properties and limitations of the G-CDHO and the Maxwell-Helmholtz-Drude models have been discussed in the context of the Kramers-Kronig relations. The dielectric parameters reported here are representative of low-stress SiO$_{\mbox{\scriptsize x}}$ membranes encountered in our fabrication and test efforts.

\noindent {\large\textbf{Funding.}} Funded by the National Aeronautics and Space Administration (NASA) under NNH12ZDA001N-APRA and the NASA Goddard Space Flight Center Internal Research and Development program.

\noindent {\large\textbf{Acknowledgment.}} The authors thank M.~A.~Quijada for helpful discussions. G.~C. thanks the Universities Space Research Association for the administration of his appointment at NASA.


\end{document}